\input harvmac
\noblackbox
%%%%%%%%%%%%%%%%%%%%%%%%%  %%%%%%%%%%%%%%%%%%%%%%%%%%%%%%%
% this set of macros is an extension of harvmac
% (9/91 or later) and its hypertex version, lanlmac.
% for info on hypertex, see
% http://xxx.lanl.gov/hypertex/
%%%%%%%%%%%%%%%%%%% %%%%%%%%%%%%%%%%%%%%%%%%%%%
% determine hypertex mode
%%%%%%%%%%%%%%%%%%%%%%% %%%%%%%%%%%%%%%%%%%%%%%
\newif\ifdraft

\catcode`\@=11
\newif\iffrontpage
%\draft
\newif\ifxxx
\xxxtrue
%\xxxfalse

\newif\ifad
\adtrue
\adfalse

\parindent0pt

\def\{{\lbrace}
\def\}{\rbrace}

\def\p{\partial}

\def\sss{\scriptscriptstyle}

 %gzero
 %gone
 %gtwo
 %gthree
 %gfour
 %gd
 %gn
 %hd
 %aone
 %atwo
 %athree
 %afour
 %afive
 %an
 %anminusone
 %Rzero
\def\Nz{\mathrel{\mathop \nabla^{\scriptscriptstyle{(0)}}}} %nablazero

\def\Box#1{\mathop{\mkern0.5\thinmuskip
           \vbox{\hrule\hbox{\vrule\hskip#1\vrule height#1 width 0pt\vrule}
           \hrule}\mkern0.5\thinmuskip}}
\def\box{\displaystyle{\Box{7pt}}}   
%
%

%\draftmode

%
%%%
\def\abstract#1{
\vskip .5in\vfil\centerline
{\bf Abstract}\penalty1000
{{\smallskip\ifx\answ\bigans\leftskip 2pc \rightskip 2pc
\else\leftskip 5pc \rightskip 5pc\fi
\noindent\abstractfont \baselineskip=12pt
{#1} \smallskip}}
\penalty-1000}

\def\sss{\scriptscriptstyle}

\vbadness=10000

\parindent10pt

%%%%%%%%%%%%%% Refernces %%%%%%%%%%%%%%%%%%%%%%%%
%
\lref\AGMOO{O.~Aharony, S.S.~Gubser, J.~Maldacena, H.~Ooguri and Y.~Oz,
``Large N field theories, string theory and gravity,''
hep-th/9905111.}
\lref\DS{S.~Deser and A.~Schwimmer,
``Geometric classification of conformal anomalies in arbitrary dimensions,''
Phys.\ Lett.\ {B309} (1993) 279,
hep-th/9302047.}
\lref\FG{C. Fefferman and R. Graham, ``Conformal Invariants'', 
Ast\`erisque, hors s\'erie, 1995, p.95.}
\lref\wittenone{E.~Witten,
``Anti-de Sitter space and holography,''
Adv.\ Theor.\ Math.\ Phys.\ {2} (1998) 253,
hep-th/9802150.}
\lref\Pen{R.~Penrose and W.~Rindler, ``Spinors and Spacetime'', 
vol 2, CUP 1986, chapter 9.}
\lref\APTY{O.~Aharony, J.~Pawelczyk, S.~Theisen and S.~Yankielowicz,
``A note on anomalies in the AdS/CFT correspondence,''
Phys.\ Rev.\ {D60} (1999) 066001,
hep-th/9901134.}
\lref\HS{M.~Henningson and K.~Skenderis,
``The holographic Weyl anomaly,''
JHEP {07} (1998) 023,
hep-th/9806087.}
\lref\BPB{L.~Bonora, P.~Pasti and M.~Bregola, ``Weyl cocycles'', 
Class. Quantum. Grav. 3 (1986) 635.}
\lref\CFL{A.~Cappelli, D.~Friedan and J.I.~Latorre,
``C theorem and spectral representation,''
Nucl.\ Phys.\ {B352} (1991) 616.}
\lref\BH{J.D.~Brown and M.~Henneaux,
``Central Charges In The Canonical Realization Of Asymptotic Symmetries: 
An Example From Three-Dimensional Gravity,''
Commun.\ Math.\ Phys.\ {104} (1986) 207.}
\lref\Strings{
http://strings99.aei-potsdam.mpg.de/cgi-bin/viewit.cgi?speaker=Theisen}
\lref\Cardy{J.L.~Cardy,
``Is There A C Theorem In Four-Dimensions?,''
Phys.\ Lett.\ {\bf B215} (1988) 749.}
\lref\BGNNO{S.~Nojiri and S.D.~Odintsov,
``On the conformal anomaly from higher derivative gravity in 
AdS/CFT  correspondence,'' hep-th/9903033;
M.~Blau, K.S.~Narain and E.~Gava,
``On subleading contributions to the AdS/CFT trace anomaly,''
JHEP {\bf 09} (1999) 018
hep-th/9904179.}
                            
\Title{\vbox{
\rightline{\vbox{\baselineskip12pt
\hbox{TAUP-2590-99}
\hbox{hep-th/yymmddd}}}}}
{Diffeomorphisms and Holographic Anomalies
\footnote{$^{\scriptscriptstyle*}$}{\sevenrm 
Partially supported by GIF, 
the German-Israeli Foundation for Scientific Research,
by the European Commission TMR programme ERBFMRX-CT96-0045, Minerva Foundation,
the Center for Basic Interactions of the Israeli Academy of Sciences
and the US-Israel Binational Science Foundation
}}
\vskip 0.3cm
\centerline{C.~Imbimbo$^a$, A.~Schwimmer$^b$, 
S.~Theisen$^c$ and S.~Yankielowicz$^d$}
\vskip 0.6cm
\centerline{$^a$ \it Dipartimento di Fisica dell' Universit\`a di Lecce,
Via Arnesano, I-73100, Lecce, Italy}
\vskip.2cm
\centerline{$^b$ \it Department of Physics of the Complex Systems, 
Weizmann Institute, Rehovot 76100, Israel} 
\vskip.2cm
\centerline{$^c$ \it Sektion Physik, Universit\"at M\"unchen, Germany}
\vskip.2cm
\centerline{$^d$ \it School of Physics and Astronomy, Beverly and Raymond
Sackler Faculty of Exact Sciences,}
\centerline{\it Tel-Aviv University, Ramat Aviv, Tel-Aviv 69978, Israel}
\vskip 0.0cm

\abstract{Using the relation between diffeomorphisms in 
the bulk and Weyl transformations on the boundary we study the 
Weyl transformation properties of the bulk metric on shell
and of the boundary action.
We obtain a universal formula for one of the classes of trace anomalies
in any even dimension in terms of the parameters of the gravity action.
}
\Date{\vbox{\hbox{\sl {November 1999}}
}}
\goodbreak

\parskip=4pt plus 15pt minus 1pt
\baselineskip=15pt plus 2pt minus 1pt

%%%%%%%%%%%%%%%%%%%%%%%%%%%%%%%%%%%%% %%%%%%%%%%%%%%%%%%%%%%%%%%
\newsec{Introduction}

The AdS/CFT correspondence offers remarkable insights into
nonperturbative phenomena in gauge theories \AGMOO.
Many of the proposed tests of the correspondence rely on the symmetry
algebras being isomorphic.

Among the tests going beyond the mapping of 
the algebraic structure the correct mapping of the trace anomalies is one
of the most impressive \wittenone,\HS,\APTY. 
On the supergravity side the correspondence involves a classical calculation:
one solves the equation of motion using the metric at the boundary as initial
condition. The action evaluated for this classical solution gives the
effective  action in terms of the boundary metric. Taking a Weyl variation 
of the effective action gives the anomalous terms. An anomaly appears in a 
classical calculation due to the apparently 
infrared logarithmically divergent terms obtained when the action is 
evaluated  with the classical solution. For the five dimensional
Einstein action with a cosmological constant the 
coefficients of the two independent trace anomaly structures match correctly
the trace anomalies of the four dimensional N=4 super Yang Mills theory
calculated in the large N limit.

In the present note we study further the structure of this correspondence.
Following Penrose \Pen\
and Brown and Henneaux \BH\ we remark that the Weyl transformations
of the boundary metric can be understood as a certain subgroup of the bulk
diffeomorphisms.
This observation allows us to derive a general transformation rule for the bulk
metric when the boundary metric is changed by a Weyl transformation.
We will call it in the following the ``PBH transformation''. The
transformation doesn't use explicitly the solution of the equation
of motion and, therefore, it is valid for a general bulk effective action
including all the stringy corrections. We analyze the action evaluated for 
this general bulk metric and we uncover some universal properties 
of the trace anomaly which follow from it. In particular the coefficient
of the Euler structure present in any dimension (the so called type A
trace anomaly) has a universal dependence on the action, being given by
the value of the action evaluated for the AdS solution. This allows us to  
calculate this particular trace anomaly for a general bulk action.
The universal structure we find has general features reminiscent of the 
descent equations for the axial anomaly and probably has implications for
the general understanding of the structure of trace anomalies,
independent of the AdS/CFT correspondence.

In Section 2 we discuss the algebraic structure of the PBH transformation.
In Section 3 we classify the terms in the effective 
action using the symmetry.
The relation between the terms in the effective 
action and trace anomalies and a general formula for the type A  trace 
anomaly are discussed in Section 4.
The general conclusions we have reached
and possible implications 
and open questions are discussed in the last Section.  

Most of the results of this paper were already presented at 
Strings 99 \Strings.

%%%%%%%%%%%%%%%%%%%%%%%%%%%%%%%%%%%%%%%%%%%%%%%%%%%%%%%%%%%%%%%
\newsec{The PBH transformation}
Consider a manifold in $d+1$ dimensions with  a boundary
which is topologically $S_d$.
Following Fefferman and Graham \FG, one can choose a set of coordinates
in which the $d+1$ dimensional metric has the form:
\eqn\ansatz{
ds^2=G_{\mu\nu} dX^\mu dX^\nu={l^2\over4}\left({d\rho\over\rho}\right)^2
+{1\over\rho}g_{ij}(x,\rho)dx^i dx^j\,.
}
Here $\mu,\nu=1,\dots,d+1$ and $i,j=1,\dots,d$. The coordinates are 
chosen such that $\rho=0$ corresponds to the boundary. 
We will assume that $g_{ij}$ is regular at $\rho=0$,
$g_{ij}(x,\rho=0)$ being the boundary metric. 

We now look for those $(d+1)$-dimensional diffeomorphisms which 
leave the form of the metric invariant. We make the ansatz
\eqn\diffeo{\eqalign{
\rho&=\rho' e^{-2\sigma(x')}\simeq \rho'(1-2\sigma(x'))\,,\cr
x^i&=x'^i+a^i(x',\rho')\,.}}
The $a^i(x',\rho')$ are infinitesimal, and are restricted by the requirement
of form invariance of the metric. We will work 
to ${\cal O}(\sigma,a^i)$. We insert  \diffeo\ in \ansatz\ and
require that the $dx' d\rho'$ components of the metric 
vanish. This gives 
\eqn\prhoa{
\partial_\rho a^i ={l^2\over2}g^{ij}\partial_j\sigma\,.}
With the boundary condition  $a^i(x,\rho\!=\!0)=0$ this integrates to
\eqn\ais{
a^i(x,\rho)={l^2\over2}\int_0^\rho d\rho'g^{ij}(x,\rho')\partial_j\sigma(x)\,.}
Performing the diffeomorphism defined by \diffeo, $g_{ij}$ will
generally transform:  
\eqn\dg{
\delta g_{ij}(x,\rho)=2\sigma(1-\rho\partial_\rho)g_{ij}(x,\rho)
+\nabla_i a_j(x,\rho)+\nabla_j a_i(x,\rho)\,.
}
The covariant derivatives are with respect to the metric $g_{ij}(x,\rho)$
where $\rho$ is considered a parameter.

The equations \diffeo, \ais, \dg,
define the BPH transformation, 
i.e.~a subgroup of bulk diffeomorphisms
which leave the metric in
the form \ansatz\ and
which on the boundary reduce
to a Weyl transformation.

We assume that $a^i(x,\rho)$ and $g_{ij}(x,\rho)$ 
have power series expansions in the vicinity of $\rho=0$, i.e.~we write 
\eqn\expansiona{
a^i(x,\rho)=\sum_{n=1}^\infty a_{\sss (n)}^i(x)\rho^n}
and
\eqn\expansiong{
g_{ij}(x,\rho)=\sum_{n=0}^\infty g_{{\sss (n)}ij}(x)\rho^n\,.}
{}For a general bulk metric the coefficients in the expansion
are arbitrary. If however, the metric is the solution of an equation of 
motion, the coefficients of the expansion are all expressible in terms
of the boundary metric $g_{{\sss(0)}ij}$. This result was
proven by Fefferman and Graham
for the equation of motion following from the Einstein action with 
cosmological term; we will assume that it holds for any action which
admits an AdS solution.
{}For $d$ an even integer, the expansion for $g$ has also logarithmic terms. 
Using dimensional regularization (i.e.~$d$ non-integer)
these terms are absent in the Fefferman--Graham case; we assume that this 
feature holds also for the general case.
The coefficients in the expansion
of $g_{ij}(x,\rho)$ are covariant tensors built from 
the boundary metric $g_{{\sss(0)}ij}(x)=g_{ij}(x,\rho=0)$,
since general covariance in $d$ dimensions is explicitly kept.
Moreover by a simple scaling argument 
$g_{\sss(n)}$ contains $2n$ derivatives with respect to the $x$
variables.
The PBH transformation implies  that the behavior of $g_{\sss(n)}$
under a Weyl transformation of the boundary metric is known.
This determines to a great extent the expressions of $g_{\sss(n)}$
in terms of covariant tensors built of $g_{{\sss(0)}ij}$. We perform
now the calculation  outlined above.

We first express the $a_{\sss (n)}$ 
in terms of the $g_{\sss (n)}$. The first few  terms are
\eqn\ais{
\eqalign{a^i_{\sss(1)}&={l^2\over 2}g_{\sss(0)}^{ij}\partial_j\sigma\,,\cr
a^i_{\sss(2)}&=-{l^2\over 4}g_{\sss(1)}^{ij}\partial_j\sigma\,,\cr
a^i_{\sss(3)}&={l^2\over 6}\bigl[ g_{\sss(1)}^{ik}g_{{\sss(1)}k}{}^j
-g_{\sss(2)}^{ij}\bigr]\partial_j\sigma\,,\cr
a^i_{\sss(4)}&={l^2\over8}\bigl[-g_{\sss(3)}^{ij}
+g_{\sss(1)}^{ik}g_{{\sss(2)}k}{}^j
+g_{\sss(2)}^{ik}g_{{\sss(1)}k}{}^j-g_{\sss(1)}{}^i{}_k 
g_{\sss(1)}{}^k{}_l g_{\sss(1)}^{lj}
\bigr]\nabla_j\sigma\,.}}

The variations  of $g_{\sss (n)}$ under a Weyl transformation of
$g_{{\sss(0)}ij}$ are easily obtained by combining \diffeo, \ais\ and
\dg\ with the expansion \expansiong. We give again just the 
first few terms:
\eqn\ges{\eqalign{
\delta g_{{\sss(0)}ij}&=2\sigma g_{{\sss(0)}ij}\,,\cr
\delta g_{{\sss(1)}ij}&=\Nz_i a_{{\sss(1)}j}+\Nz_j a_{{\sss(1)}i}\,,\cr
\delta g_{{\sss(2)}ij}&=-2\sigma g_{{\sss(2)}ij}+a_{(1)}^k\Nz_k g_{{\sss(1)}ij}
+\Nz_i a_{{\sss(2)}j}+\Nz_j a_{{\sss(2)}i}
+g_{{\sss(1)}ik}\Nz_j a_{\sss(1)}^k+g_{{\sss(1)}jk}\Nz_i a_{\sss(1)}^k\,.}}
The covariant derivative ${\displaystyle \Nz_i}$ is w.r.t. $g_{\sss(0)}$.
These equations can be integrated w.r.t. $\sigma$. For the first 
two non-trivial terms in the expansion \expansiong\ we find
\eqn\firstges{\eqalign{
g_{{\sss(1)}ij}&=
{l^2\over d-2}\Bigl[R_{ij}-{1\over 2(d-1)}R g_{{\sss(0)}ij}\Bigr]\,,\cr
g_{{\sss(2)}ij}&=c_1 l^4 C_{ijkl}C^{ijkl}g_{{\sss(0)}ij}
+c_2 l^4 C_{iklm}C_{j}{}^{klm}\cr
&+{l^4\over d-4}\Bigl\lbrace -{1\over 8(d-1)}\nabla_i\nabla_j R
+{1\over 4(d-2)}\nabla^2 R_{ij}
-{1\over 8(d-1)(d-2)}\nabla^2 R g_{{\sss(0)}ij}\cr
&-{1\over 2(d-2)}R^{kl}R_{ikjl}
+{d-4\over 2(d-2)^2}R_i{}^k R_{jk}
+{1\over (d-1)(d-2)^2}R R_{ij}\cr
&+{1\over 4(d-2)^2}R^{kl}R_{kl}g_{{\sss(0)}ij}
-{3d\over 16(d-1)^2 (d-2)^2}R^2 g_{{\sss(0)}ij}\Bigr\rbrace\,.}}
Here the curvature $R$, the Weyl--tensor in $d$ dimensions $C$ and
the covariant derivative $\nabla$ are all those of the boundary metric 
$g_{\sss(0)}$.\foot{We use the following curvature conventions:
$[\nabla_i,\nabla_j]V_k=-R_{ijk}{}^l V_l$ and $R_{ij}=R_{ikj}{}^k$.}

Starting from $g_{\sss(2)}$, there are solutions to the homogeneous 
equations, i.e.~curvature invariants which transform 
homogeneously under Weyl transformations.  
E.g.~for $g_{\sss(2)}$ there are two free parameters. 
Of course we have  complete agreement with the explicit calculation
of ref.\HS.

We stress that the above expressions do not assume any specific
form of the action. The dependence on the action enters only through
the arbitrary coefficients of the homogeneous terms. Our goal 
will be to extract the universal (i.e.~independent on the arbitrary
coefficients) information about the trace anomalies.

\newsec{The PBH transformation and the effective boundary action}

We study  now the  implications of the symmetry defined above
for a general gravitational action $S$ which is invariant under $d+1$ 
dimensional diffeomorphisms: 
\eqn\generalac{
S={1\over 2\kappa_{d+1}^2}\int d^{d+1}X\sqrt{G}f(R(G))\,,}
where $f$ is a local function of the curvature and its covariant derivatives.
For the application we have in mind, we must choose $f(R)$ such that
$AdS_{d+1}$ with radius  $l$ 
is a solution of the equations of motion.
If we insert \ansatz\ into \generalac\ we get an expression of the form
\eqn\action{
2\kappa_{d+1}^2 S={l\over2}\int d\rho\, d^d x\,\rho^{-{d\over2}-1}
\sqrt{g_{\sss (0)}(x)}b(x,\rho)\,,}
where the specific form of $b(x,\rho)$ does depend on $f(R)$.
 
Since the integrand is a scalar under diffeomorphisms, the action
satisfies
\eqn\acttrans{
2\kappa^2_{d+1}S=
{l\over2}\int d\rho\,d^dx\,\rho^{-{d\over2}-1}\sqrt{g_{\sss(0)}(x)}\,b(x,\rho)
={l\over2}\int d\rho'\,d^dx'\,\rho'^{-{d\over2}-1}
\sqrt{g'_{\sss(0)}(x')}\,b'(x',\rho')\,.}
On a solution of the equation of motion $b(x,\rho)$ becomes a functional
of $ g_{\sss (0)}$ and therefore using \acttrans\
for the diffeomorphisms \diffeo\ we obtain its behavior 
under a Weyl transformation:
\eqn\btransform{
\delta b(x,\rho)=b'(x,\rho)-b(x,\rho)
=-2\sigma(x)\,\rho\,\partial_\rho b(x,\rho)
+\Nz_i\bigl(b(x,\rho)a^i(x,\rho)\bigr)\,.}
Using the explicit expression \ais, it is now straightforward to show that 
$b(x,\rho)$ satisfies the WZ consistency condition:
\eqn\WZ{
\int d^dx\sqrt{g_{(0)}}\left(\sigma_1 \delta_{\sigma_2}b
-\sigma_2 \delta_{\sigma_1}b\right)=0\,.}
We will see below that $b(x,\rho)$ may be considered as a generating 
function for Weyl anomalies in all (even) dimensions.

In order to use explicitly the Weyl transformation properties
we expand the generating function $b(x,\rho)$ in a power 
series in $\rho$: 
\eqn\bexpansion{
b(x,\rho)=\sum_{n=0}^\infty b_n(x)\rho^n\,.}
Using \bexpansion\ the $\rho$ integrations can be explicitly performed
and the boundary action $S$ expressed in terms of the $b_n$:
\eqn\missing{\kappa_{d+1}^2 S=\sum_{n=0}^\infty {l\over 2n-d}
\int  d^d x \sqrt{g_{\sss (0)}(x)}b_n(x)\,.}
For a $ b(x,\rho)$ evaluated on a solution $g(x,\rho)$ the coefficients
$b_n(x)$ become scalar local functionals of $g_{\sss(0)}$.
The BPH transformation
\btransform\ gives the transformation of $b_n(x)$ under a Weyl transformation:
\eqn\deltabgen{
\delta b_n=-2n\sigma b_n+ \nabla_i\Bigl(\sum_{m=0}^{n-1}b_m a_{\sss (n-m)}^i
\Bigr)\,.}
Explicitly, for the first few coefficients, we find
\eqn\deltabn{\eqalign{
\delta b_0&=0\,,\cr
\delta b_1&=-2\sigma b_1+{l^2\over2}b_0 \box\sigma\,,\cr
\delta b_2&=-4\sigma b_2-{b_0 l^4\over 4(d-2)}
\left[R^{ij}\nabla_i\nabla_j\sigma-{1\over2}R\box\sigma\right]\,.\cr
}}
$\delta b_3$, which is a rather unwieldy expression, contains the two 
arbitrary parameters $c_1$ and $c_2$ in $g_{{\sss(2)}ij}$.

The Weyl variation determines to a certain extent the dependence 
of $b_n(x)$  on $g_{\sss(0)}$.
We will argue in the next section that local expressions for $b_n$ which
satisfy \deltabn\ can always be found.
We give here again only the first few explicit expressions:

\eqn\solb{\eqalign{
b_0&={\rm const}\,,\cr
b_1&=b_0{l^2\over 4(d-1)}R\,,\cr
b_2&={l^4 b_0\over 32(d-2)(d-3)}E_4+c l^4 C_{ijkl}C^{ijkl}\,,\cr}}
where as before $C_{ijkl}$ is the Weyl-tensor corresponding to 
$g_{\sss(0)}$ in $d$ dimensions 
and $E_{2n}$ is the 
Euler density, which is defined in $d=2n$ dimensions as
\eqn\eulerdensity{
E_{2n}={1\over 2^n}\,R_{i_1 j_1 k_1 l_1}\cdots R_{i_n j_n k_n l_n}
\epsilon^{i_1 j_1 i_2 j_2\dots i_n j_n}\epsilon^{k_1 l_1 \dots k_n l_n}\,.}
Once the product of the epsilon-tensors has been 
expressed in terms of products of the metric, this expression 
makes sense in any dimension.

Again, starting from the second coefficient $b_2$ there is an increasing
number of solutions to the homogeneous equations which enter with 
free parameters. These arbitrary coefficients and those appearing in the 
integration of $\delta g_{\sss (n)}$ will appear in the expressions
for $\delta b_{\sss (n)}$ for $n>2$. \foot{For instance, 
the homogeneous equation $\delta b_3=-6\sigma b_3$ has three solutions. 
In addition, $b_3$ depends on the two parameters
already present in the inhomogeneous piece of 
$\delta b_3$.}
In the next section we will discuss
the various constraints on these coefficients following 
from their relation to trace anomalies.

The coefficients $b_n$ depend on the action through the constant $b_0$ 
and through the constants which enter through the solutions to the 
homogeneous equations.
For $b_0$ we can obtain a general formula. Indeed, if we
write the action \generalac\ in the form
\eqn\genact{
2\kappa_{d+1}^2 S=\int d\rho\, d^d\! x\sqrt{G}f(R(G))=
{l\over2}\int d\rho\, d^d\! x\rho^{-{d\over2}-1}\sqrt{g}f(R(G))}
then 
\eqn\bzero{
b_0=l f(R)|_{\rho=0}=l f(AdS)\,,}
where in the last expression $f$ is to be evaluated at
the action of $AdS_{d+1}$ space with 
\eqn\RAdS{
R_{\mu\nu\rho\sigma}={1\over l^2}(G_{\mu\rho}G_{\nu\sigma}
-G_{\mu\sigma}G_{\nu\rho})\,.}
To see this one simply has to realize that $\sqrt{g}f(R)$ has an expansion in 
positive powers of $\rho$ and that only the most singular (in $\rho$) 
contributions of $R_{\mu\nu\rho\sigma}$ contribute to $b_0$. 
It is straightforward to show that this contribution is as given in \RAdS.

We see, therefore, that the effective action of the boundary metric has a
structure determined to a large extent by the action of the Weyl 
transformations which at their turn are fixed by the PBH transformation.
The specific form of the bulk action manifests itself through a set of 
constants: $b_0$ which has the universal structure given by \bzero\
and the coefficients of the terms involving the Weyl tensors in
\deltabgen\ and \solb.

We remark that the general expressions \solb, being based on symmetry
considerations, don't assume that the action which gives the equation 
of motion which determines $g_{ij}(x,\rho)$ and the action
which determines $b$ should be the same, provided both are 
diffeoinvariant.
The explicit expression \bzero\ for $b_0$ needs of course
the identity of the two actions.

In the next section we will exploit the relation between the effective
action and trace anomalies in order to constrain the ambiguities on one hand
and to extract universal properties of the trace anomalies from the effective action, on the other.

\newsec{Trace anomalies and the effective action}

We start this section with a brief summary of the general structure
of trace anomalies, c.f. \BPB,\CFL,\DS, 
and the appearance of trace anomalies in boundary actions \HS.

The trace anomalies can be characterized by the anomalous Weyl variation
of a diffeoinvariant effective action depending on a
metric in $d=2n $ dimensions. This action reflects  the properties 
of a conformally invariant matter theory and it is obtained, in principle,
by coupling the matter to a classical background metric and integrating
the matter fields out. 

In general the effective action is a nonlocal functional of the metric.
Its Weyl variation, however, representing an anomaly is necessarily local.
The second variation of the action, i.e.~the first variation of the anomaly,
is a  symmetric functional of the two infinitesimal Weyl parameters. This
integrability condition is the  Wess -Zumino condition the anomaly
must satisfy.

The above stated  structure offers a general classification
of trace anomalies in any even dimension :

a) There is always the type A anomaly whose expression is the 
Euler characteristic in the respective dimension. This type doesn't reflect 
any real logarithmic ultraviolet divergence. In dimensional regularization
however, out of the term in the effective action giving rise to it,
a piece having a simple pole in $2n-d$  
multiplying the  Euler characteristic, can be separated, giving a clear
signature for this type of anomaly.

b) There is an increasing (with dimension) number of type B anomalies:
they originate in true logarithmic divergences in  correlators
of lower order than the anomalous ones. Their
signature is an expression which is $exactly$ Weyl invariant. The simplest
way to construct Weyl invariant expressions is through the contraction
of Weyl tensors but there are invariants starting with covariant derivatives
of Weyl tensors, etc. In dimensional regularization, these terms
appear directly in the effective action multiplying poles in $2n-d$.

In addition there  are cohomologically trivial local expressions, 
i.e.~obtained by Weyl variations of local, Weyl non-invariant, pieces in the 
effective action. These terms, representing arbitrary real parts (subtraction
constants) in the underlying matter theory, cannot contain
any dynamical information and their explicit form depends on
the regularization chosen.  

Going now back to \missing\ we observe that $b_n$ appears multiplying
a pole in $2n-d$: therefore  $b_n$ is a trace anomaly in $d=2n$ dimensions, 
i.e.~it could
be expanded as a linear combination of 
a type A, the type B appropriate to the dimension and cohomologically
trivial terms. Indeed, as  mentioned in the previous section, the $b_n$'s
satisfy the WZ condition, eq.\WZ. 

A first conclusion we could reach is that eq.\deltabgen\
always  have a solution in terms of local expressions which are
linear combinations of type A, type B and cohomologically trivial. Of course
this puts constraints on the terms which could appear on the r.h.s.
of \deltabgen. 

We will be more interested, however, in the way we could extract universal
information about trace anomalies from the effective action. The anomaly
of type A is a clear candidate for this type of information. Indeed, as 
we remarked before, terms involving contractions of Weyl tensors
(i.e.~type B) appear with arbitrary coefficients in $b_n$; therefore
we couldn't expect any simple, general expression for type B.

In order to isolate the general expression we are looking for in type A,
we use the following simple observation: an exactly Weyl invariant expression
vanishes identically for a metric which is conformally flat, 
i.e.~which has the form:
\eqn\confflat{g_{\sss(0){ij}}=\exp(2\phi)\delta_{ij}\,.}
Therefore the terms with arbitrary coefficients in the various
recursion relations (``the homogeneous terms'') will disappear.
As a consequence, the unambiguous solution of
the equations for a metric of the form
\confflat\ will give us the information about the surviving type A
anomaly. 

We will expand the solutions in powers of $\phi$ defined
in \confflat. In particular the Euler characteristic  
in $d=2n$ starts with the $n$-th power of $\phi$:
\eqn\euler{\eqalign{
E_{2n}&\equiv {1\over 2^n}R_{i_1 j_1 k_1 l_1}\dots R_{i_n j_n k_n l_n}
\epsilon^{i_1 j_1\dots i_n j_n}\epsilon^{k_1 l_1\dots k_n l_n}\cr
&=n!\,2^n\,(\p_{i_1}\p_{j_1}\phi)\cdots
(\p_{i_n}\p_{j_n}\phi)\delta^{i_1\dots i_n}_{j_1\dots j_n}
+{\cal O}(\phi^{n+1})\,.}}

Now, our main problem is to isolate the contribution of the
type A anomaly (proportional to the Euler characteristic) from the
cohomologically trivial pieces. We rewrite the basic recursive relation
\deltabgen\ in an integrated form by multiplying it with an infinitesimal
Weyl parameter $\sigma_1$ the variation being with a new Weyl
parameter $\sigma_2$:
\eqn\crucial{\int  d^dx\sqrt{g_{\sss(0)}(x)}\,\sigma_1\delta_2 b_n =
-2 n\int  d^dx\sqrt{g_{\sss(0)}(x)}\,\sigma_1
\sigma_2 b_n -\int  d^d x\sqrt{g_{\sss(0)}(x)}\, \nabla_i\sigma_1
\sum_{m=0}^{n-1}b_m a_{\sss (n-m)}^i\,,}
where $a^i$ contains $\sigma_2$.
{}For a metric of  the form \confflat\ and expanding in powers of 
$\phi$ it is easy to see that $b_n$ starts with a term containing
$n$ powers of $\phi$ and that $a^i$ starts with terms as follows:
\eqn\as{\eqalign{
a^i_{\sss(1)}&={l^2\over2}\p_i\sigma+{\cal O}(\phi)\,,\cr
a^i_{\sss(2)}&=-{l^4\over4}(\p_i\p_j\phi)\p_j\sigma+{\cal O}(\phi^2)\,,\cr
a^i_{\sss(3)}&={l^6\over8}(\p_i\p_k\phi)(\p_k\p_j\phi)\p_j\sigma
+{\cal O}(\phi^3)\,.}}
Using induction it is straightforward to prove that \as\ generalizes
as expected, i.e. that the term of order $n$ has the form:
\eqn\ass{
a^i_{\sss(n)}=(-)^{n+1}{l^{2n}\over 2^n}[(\partial^2\phi)^{n-1}]_{ij}\,
\partial_j\sigma+{\cal O}(\phi^n)\,.}
Now, in an expansion in $\phi$ of $b_n$ the cohomologically trivial
pieces come from the Weyl variation of a local expression containing $n+1$
powers of $\phi$. The variation replaces each field $\phi$ at its turn
by $\sigma_1$. As a consequence the result will be  symmetric under the 
interchange of each $\phi$ with $\sigma_1$. The Euler characteristic
multiplied with $\sigma_1$ doesn't share this symmetry property. 
Therefore,  
if we antisymmetrize eq.\crucial, i.e.~if from the equation multiplied 
with $n$ we subtract  $n$ terms where $\sigma_1$ was interchanged
with each factor $\phi$ and with $\sigma_2$ respectively, we will be left
on the l.h.s. with the contribution of the Euler term only.
\foot{The term with $\sigma_1$ interchanged with $\sigma_2$ automatically
cancels the original term, this being the content of the WZ condition;
we could have left it out and changed the normalization to $n-1$}

We can now specialize even further and pick one particular term in the 
Euler characteristic e.g.~the contribution 
$n!2^n (\box\phi)^n $. This term will give 
a contribution $\sigma_1 \box \sigma_2 (\box \phi)^{n-1}$ to the integrand
of the l.h.s. of \crucial, antisymmetrized of course, and with
a numerical coefficient following from our normalizations.

We can try to match this term with terms on the r.h.s. of \crucial.
{}From the explicit form of $a^i$ we see that all terms on the r.h.s.
except the last, $(n-1)\,-\,$th one in the sum,
will have two partial derivatives not contracted
into a $\box$. In the last term however, we have after a partial integration
a contribution to the integrand of the form:
$\sigma_1 \box\sigma_2 b_{n-1} $  antisymmetrized. Again the 
antisymmetrization projects out from $b_{n-1}$ all contributions but
the Euler one in $(n-1)^{\rm th}$ order in $\phi$. 
In particular the $(\box\phi)^{n-1}$
contribution to $E_{2(n-1)}$ matches nicely the l.h.s. 

We obtained, therefore a recursion 
relation between the Euler contributions alone, in different orders.
Reintroducing the numerical factors we left out in the previous argument
the recursion can be easily solved and we obtain for the Euler contribution
the general formula:
\eqn\bn{
b_n={ l^{2n}b_0\over 2^{2n}(n!)^2}E_{2n}\!
+\!\hbox{cohomologically trivial terms}\,.}

Using  the  expression \bzero\ for $b_0$ evaluated for $d=2n$
we have now a general formula for the type A trace anomaly
in any even dimension  corresponding
to a given gravity action which admits an AdS solution. We have
to think about the gravity action given abstractly as a polynomial
in the curvature without specifying the dimension. The specific form
of the action enters in determining the radius of the AdS solution and then
in the evaluation of the action on the solution, to give $b_0$.
All the factors depending on the dimension $d=2n$ in the expression
for the coefficient of the type A anomaly are then explicit.

As we remarked already in the previous section the symmetry considerations
do not require that the action which determines $g$ and the one
which determines $b$ should be the same. The universal dependence on
the dimension appears, however, only if the two actions are identified.

\newsec{Discussion}

Through the PBH argument the Weyl transformations of the effective boundary
theory can be understood to follow from the diffeomorphisms of the bulk 
theory. The dependence on the ``odd'' variable $\rho$ in the bulk gets
translated into the dependence on the dimension, $b(x,\rho)$ playing
the r\^ole of a generating function for the type A trace anomalies in various
dimensions.

The final result is very similar in spirit to the ``elliptic genus''
which gives, through the ``descent equations'', a unifying expression for 
the chiral anomaly in various dimensions. 
The elliptic genus gives an expression for the chiral anomaly
of gauge theories with chiral fermions
in various dimensions which have the same gauge group
and representation content of the fermions.
In this analogy there is, however,
a part we really don't understand, i.e. which is the common feature
of the conformal field theories in various dimensions whose trace anomalies
are represented by the same gravitational action. It is tempting
to conjecture that general properties like e.g. the ``flow'' as measured by 
the anomaly coefficient, are common in these theories. This would suggest
that the type A anomaly coefficient in $d=4$  related to $c$ in $d=2$
gives the right generalization of the $c$-theorem as first proposed
by Cardy \Cardy.

Beginning with $b_3$ the Euler characteristic appears accompanied by
cohomologically trivial terms which are determined by the recursion relations.
We don't understand the physical significance, if any, of these terms.
More generally, it would be interesting to study the exact mathematical
structure of the ``cohomological'' equations \ais, \ges\ and \deltabn.

Finally, the special r\^ole the type A anomaly plays in the discussion
suggests that it should be singled out for the understanding 
of the non-leading $1/N$ corrections in the AdS/CFT correspondence
which are still not under control \BGNNO. 
We postpone the discussion of this problem to a later publication.

\listrefs
\end